\documentstyle[multicol,aps,prl,epsf,rotate]{revtex}

\date{\today}
\draft
\tighten
\newcommand{\be}{\begin{equation}}
\newcommand{\ee}{\end{equation}}
\newcommand{\bea}{\begin{eqnarray}}
\newcommand{\eea}{\end{eqnarray}}
\begin{document}
\def\sqr#1#2{{\vcenter{\hrule height.3pt
      \hbox{\vrule width.3pt height#2pt  \kern#1pt
         \vrule width.3pt}  \hrule height.3pt}}}
\def\square{\mathchoice{\sqr67\,}{\sqr67\,}\sqr{3}{3.5}\sqr{3}{3.5}}
\def\today{\ifcase\month\or
  January\or February\or March\or April\or May\or June\or July\or
  August\or September\or October\or November\or December\fi
  \space\number\day, \number\year}

\def\Bbb{\bf}
\topmargin=-0.2in


\newcommand{\ww}{\mbox{\tiny $\wedge$}}
\newcommand{\pp}{\partial}

\title{Graviton and Radion Production at LHC: From pp and PbPb Collisions}
\author{Gouranga C. Nayak}
\address{T-8, Theoretical Division, Los Alamos National Laboratory,
Los Alamos, NM 87545, USA }
\maketitle

\begin{abstract}
If the Planck scale is around a TeV in theories with large extra dimension, 
then gravitons should be
produced in high energy hadronic collisions at the LHC. In this paper we
compute the direct graviton production cross section in a pp and
PbPb collision at LHC at $\sqrt s^{NN}$= 14 TeV and 5.5 TeV respectively.
The graviton production cross section in a lead-lead collision
is enhanced in comparison to a pp collision at LHC
depending upon the minimum transverse energy of the jet 
$E_{T,jet}^{min}$ above which the graviton production cross section is
computed. For two extra dimensions and above $E_{T,jet}^{min}$ = 100, 500, 
1000, 1500 GeV 
the ratio of graviton production cross sections in a PbPb collision
to that in a pp collision is found to be
2220, 400, 43, 2. For four extra dimensions 
this ratio is found to be 470, 100, 10, 0.5. In the Randall-Sundrum
model the radion production
cross section is also found to be enhanced in a lead-lead collision over that
in a pp collision at LHC. The ratio of the radion production cross sections
in a PbPb collision to that in a pp collision at LHC is found to be
1650, 460, 180, 67, 23, 1 for radion of masses 100, 500, 1000, 1500, 2000, 
3000 GeV respectively.

\end{abstract}


\begin{multicols} {2}

\section{Introduction}
In the large extra dimension theories \cite{ADD}
it has been proposed that while the
standard model particles live in the usual 3+1 dimensional space, gravity
can propagate in higher dimensional space. Furthermore it has been 
proposed that the size of the compactified extra dimensions can be very
large. If this scenario is true then quantum gravity
effects can be realized at scale much lower than the Planck mass, 
possibly at the TeV scale. This is a very interesting proposal because 
future collider experiments can probe the effects of the quantum gravity
at the TeV scale. An important consequence of these ideas is that we should
be able to produce gravitons in future collider experiments. When the
extra dimension is compactified on a circle (for example) with size R 
the gravitons propagating in the extra dimensions appear as a tower
of massive states with almost continuous masses \cite{wells}. The coupling
of these gravitons with ordinary matter is determined by the 
gravitational interaction. Naively one would expect that the 
interaction of the graviton goes as the inverse powers of the 
Planck mass and hence very suppressed. However, the large 
phase-space of the compactfied space enters into the scattering 
phase-space and the dependence of the Planck mass exactly cancels.
Finally one has scattering cross sections which depend on the 
TeV scale Planck mass and hence can be measured at the collider experiments.

The interaction of graviton with the partons is 
obtained from the d-dimensional graviton interaction
lagrangian:
\bea
{\cal L}~=~ -\frac{1}{M_P} G^{(n)}_{\mu \nu} T^{\mu \nu}.
\eea
In the above equation $T^{\mu \nu}$ is the Yang-Mills energy momentum
tensor with quark parts taken into account and $G^{(n)}_{\mu \nu}$ is the
graviton with Kaluza-Klein mode $n$ which arises because of the
compactified extra dimension. $M_P$ is the TeV scale Planck mass. The Feynman
rules for for quark, gluon interaction with the graviton can be
obtained from the above equation. For direct graviton production
in the hadronic collisions at LHC we consider the $ 2 \rightarrow 2$
($ij \rightarrow kG$) partonic level collisions where $i,j,k=q(\bar q) g$
and $G$ is the graviton.
The final state quark or gluon (jet) is accompanied by a graviton
$G$. The graviton has Kaluza-Klein mode $n$ and we must sum over n.
However, for not too large extra dimensions 
the width of the mass splitting is very small
and the gravitons are considered as having continuous mass distribution
which has to be integrated to obtain inclusive graviton-jet production
cross section. As the final state graviton interacts 
very weakly with the ordinary matter (detector), the emission gives
rise to missing transverse energy as signatures of graviton production.
 
In the 5-dimensional Randall-Sundrum (RS) model \cite{RS,RS2} the extra 
dimension
is a single $S^1/Z_2$ orbifold, in which two three-branes of opposite tensions
reside at the two fixed points ($\phi$=0 and $\phi = \pi$) 
and a cosmological constant in the bulk serves as the source 
of five-dimensional gravity. In this case the extra dimension is small, but
the background metric is not flat. 
The scalar massless radion field $b(x)$ has been introduced
by considering the metrics of the form:
\bea
ds^2~=~e^{-2m_0|y|b(x)} g_{\mu \nu}(x) dx^{\mu} dx^{\nu} -b^2(x)dy^2  
\eea
where $y$ is the fifth coordinate and $g_{\mu \nu}(x)$ is the four-dimensional
graviton. The radius of the extra dimension is associated with the vacuum
expectation value of the massless four dimensional radion field. As this
modulus field has zero potential the size of the extra dimension is not 
determined by the dynamics of the model. A mechanism of radion stabilization
was proposed in \cite{radion} which arises classically from the presence
of a bulk scalar field propagating in the background solution of the metric
which generates the required potential to stabilize the radion field 
\cite{john}.
The minimum of the potential can be arranged to solve the hierarchy problem
without fine tuning of parameters. As a consequence the radion gets a
mass which is likely to be lighter than the Kaluza-Klein modes of any 
bulk field \cite{csaba,john1,john2}. 
The mass of the radion is $\sim$ TeV and the detection of this
radion at collider experiments will be the first signature of RS model
and stabilization mechanism.  

Based on these ideas 
there have been a number of papers which calculate
the direct graviton \cite{wells,wells1} and radion \cite{rad,park,cheung,othrs} 
production in pp collisions at $\sqrt s$ = 14 TeV at LHC and at other
experiments such as in p$\bar p$ collisions at Tevatron and at $e^+e^-$
colliders. The aim of this paper is to compute the
graviton and radion production cross section in a lead-lead collision
at LHC where it is expected that the graviton and small mass radion
production cross section should be much larger than that in a pp 
collision at LHC. This is similar to the black hole \cite{bnk1,bnk2,andr} 
and string ball \cite{stk} production at LHC where it is shown that the
small mass black hole and string ball production cross section is much
higher in a PbPb collision \cite{andr} than in a pp collision at LHC. At present
Relativistic Heavy Ion Collider (RHIC) at BNL collide two gold nuclei
(to produce quark-gluon plasma \cite{nay}) at $\sqrt s^{NN}$= 200 GeV,
which is insufficient to probe any TeV scale physics, such as production
of graviton and radion at TeV scale Planck mass. In order to create 
TeV scale particles the center of mass energy of two partons must be
larger than TeV which is not the case at RHIC. The future collider at
LHC will collide two Lead nuclei at $\sqrt s^{NN}$ = 5.5 TeV in addition
to pp collisions at $\sqrt s$ = 14 TeV. As Lead nuclei at high energy
consists of many more partons than proton, the total cross section
is expected to be much larger in a PbPb collision than in a pp collision
at the same center of mass energy. 
The larger production cross section in case of a PbPb collision comes
from larger number of partonic collisions in comparison to a pp collision.
The total center of mass of energy
of the PbPb system is 5.5 $\times$ 208 = 1144 TeV which is much 
larger than the 14 TeV total center of mass energy of the pp system.
However, as pp collision is at $\sqrt s^{NN}$= 14 TeV and
PbPb collision is at $\sqrt s^{NN}$ = 5.5 TeV the total cross section
in a PbPb collision is not equal to $\sim$ 208 $\times$ 208 times 
larger than that of the cross section in a pp collision. The question
of shadowing and saturation of parton distribution function
need to be addressed in case of large nuclei collisions at ultra high energy
such as in the case of PbPb collisions at LHC. In this paper
we will present detailed calculation of the direct graviton and radion
production cross section in PbPb collisions and will make comparison 
with that in pp collisions at LHC.

The paper is organized as follows. In section II and III we present 
the partonic level calculations of the graviton and radion production 
cross sections in hadronic collisions. In section IV we 
present the results for pp, PbPb collisions at LHC and
conclude in section V.

\section{Graviton Production at LHC}

The differential cross section for the production of direct 
graviton plus a single
jet in hadron-hadron (A-B) collisions at zero impact parameter
is given by \cite{wells,wells1}:
\bea
&&\frac{d^4\sigma_{AB \rightarrow jet +G}(s)}{
dm_G~dp_{T_{jet}}~dy_{jet}~dy_G}~=~2p_T~ m_G^{d-1}\frac{\Omega_{d-1} 
{\bar{M}}_P^2}{M_P^{d+2}}\nonumber \\
&&\sum_{i,j} F_{i/A}(x_1,Q^2) F_{j/B}(x_2,Q^2)
\frac{d\sigma_{ij \rightarrow kG}(\hat s, \hat t)}{d \hat t}.
\label{gr}
\eea
Here $F_{i/A}(x_1,Q^2)$($F_{j/B}(x_2,Q^2)$) 
are the parton structure functions of the hadorn $A$($B$) with
$x_1 (x_2)$ being the longitudinal momentum fraction of the parton 
inside the hadron A(B). $Q$ is the momentum scale at which the structure
function is evaluated, $m_G$ is the
graviton mass, $M_P$ is the TeV scale Planck mass, $\bar{M}_P$ is the
Planck mass and $p_T$ is the
transverse momentum of the graviton (or jet). $d$ is the number of
extra dimension and $\Omega_{d-1}$ is the surface of the unit-radius
sphere in $d$ dimensions. 
$\frac{d\sigma_{ij \rightarrow kG}(\hat s, \hat t)}{d \hat t}$ is the
partonic level differential scattering cross section for the process
$ij \rightarrow kG$ where $i,j,k=q,\bar q, g$ and $G$ is the graviton.
The partonic level differential cross sections for the processes:
$ij \rightarrow kG$ are given by:
\bea
&&\frac{d\sigma_{q\bar q \rightarrow gG}(\hat s, \hat t)}{d \hat t}~=~
\frac{\alpha_s}{36 {\bar M}_P^2}~\frac{\hat s}{
\hat t(m_G^2-\hat s- \hat t)}~
[\frac{-4\hat t}{\hat s}(1+ \frac{\hat t}{\hat s})(1+2 \frac{\hat t}{\hat s}
\nonumber \\
&&+2 \frac{\hat t^2}{\hat s^2})+ \frac{m_G^2}{\hat s}(1+6 \frac{\hat t}{\hat s}
+18 \frac{\hat t^2}{\hat s^2}+16 \frac{\hat t^3}{\hat s^3})-6 
\frac{m_G^4 \hat t}{\hat s^3} (1+2 \frac{\hat t}{\hat s})+ \nonumber \\
&&\frac{m_G^6}{\hat s^3} (1+\frac{4\hat t}{\hat s})],
\eea
for the $q\bar q \rightarrow gG$ process:
\bea
&&\frac{d\sigma_{q (\bar q) g \rightarrow q(\bar q)G}
(\hat s, \hat t)}{d \hat t}~=~
\frac{\alpha_s}{96 {\bar M}_P^2}~\frac{\hat s}{ \hat t(m_G^2-\hat s- \hat t)}~
[\frac{-4\hat t}{\hat s}(1+ \frac{\hat t^2}{\hat s^2}) + \nonumber \\
&&\frac{m_G^2}{\hat s}(1+ \frac{\hat t}{\hat s})(1+8\frac{\hat t}{\hat s}
+\frac{\hat t^2}{\hat s^2})-3 \frac{m_G^4 }{\hat s^2}(1+4 \frac{\hat t}{\hat s}
+\frac{\hat t^2}{\hat s^2})+ \nonumber \\
&&4\frac{m_G^6}{\hat s^3} (1+\frac{\hat t}{\hat s})-2 \frac{m_G^8}{\hat s^4}],
\eea
for the $q(\bar q)g \rightarrow q(\bar q) G$ process and:
\bea
&&\frac{d\sigma_{gg \rightarrow gG}(\hat s, \hat t)}{d \hat t}~=~
\frac{3\alpha_s}{16 {\bar M}_P^2}~\frac{\hat s}{ \hat t(m_G^2-\hat s- \hat t)}~
[1+2\frac{\hat t}{\hat s}+3\frac{\hat t^2}{\hat s^2}+2
\frac{\hat t^3}{\hat s^3} \nonumber \\
&&+\frac{\hat t^4}{\hat s^4}
-2\frac{m_G^2}{\hat s} 
(1+\frac{\hat t^3}{\hat s^3})+3 \frac{m_G^4 }{\hat s^2}(1+ 
\frac{\hat t^2}{\hat s^2})-2\frac{m_G^6}{\hat s^3}(1+\frac{\hat t}{\hat s})
+ \frac{m_G^8}{\hat s^4}]
\eea
for the $gg \rightarrow g G$ process.
In the above expression: $\hat s~=~(p_i+p_j)^2$ and $\hat t~=~(p_i-p_G)^2$.

As the graviton mass splitting is very small in the flat extra dimension
models the mass $m_G$ of the graviton is treated as continuous variable
rather than the discrete mass level in the extra dimension. Therefore the 
mass $m_G$ is integrated out in eq. (\ref{gr}) in order to obtain
inclusive cross section in hadronic collisions at LHC. As the total
cross section is infrared divergent we compute the total inclusive
cross section above certain transverse momentum of the jet
$p_{T, jet}~=~E_{T,jet}$.
The factorization and re-normalization scale appearing in $F_{i/A}(x_1,Q^2)$
($F_{j/B}(x_2,Q^2)$) and $\alpha_s(Q^2)$ respectively,
are taken to be at $Q^2~=~M_T^2~=~m_G^2+p_T^2$.

\section{Radion Production at LHC}

The interaction of the radion ($\phi$) with the standard model particles
is given by: 
\bea
{\cal L}~=~\frac{\phi}{\lambda_{\phi}} T^\mu_\mu(SM)
\eea
where the strength of the coupling of the radion to the standard model 
particles $\frac{1}{\lambda_{\phi}}$ is order of $\sim$ 1/TeV. In the 
above equation $T^\mu_\mu (SM)$ is the trace of the SM energy-momentum
tensor: 
\bea
&&T^\mu_\mu~=~\sum_{f} m_f \bar f f -2m_W^2 W^+_\mu W^{-\mu} \nonumber \\
&& -m_Z^2 Z_\mu Z^\mu
+(2m_h^2 h^2 -\partial_\mu h \partial^\mu h)+...
\eea
where $f$ is fermion and $h$ is Higgs boson. For the coupling of the radion
to a pair of gluons comes from two parts: 1) contributions from one-loop
diagrams with the top quark (and $W$) in the loop and 2) from the trace
anomaly \cite{trace}: 
\bea
T^\mu_\mu (SM)^{anom}~=~\sum_a \frac{\beta_{QCD}}{2g_s} F^a_{\mu \nu} F^{a
\mu \nu}.
\eea
Including the one-loop top quark and trace anomaly contributions
the effective coupling of a radion and two gluons $\phi gg$ is given by:
\bea
&&\frac{i \delta^{ab} \alpha_s}{2\pi \lambda_{\phi}}[7+
\frac{4 M_t^2}{2 p_1 \cdot p_2}[1+(1- \frac{4 M_t^2}{2 p_1 \cdot p_2})
g(\frac{4M_t^2}{(p_1+p_2)^2})]] \nonumber \\
&& \times [p_1 \cdot p_2 g_{\mu \nu} -{p_2}_\mu {p_1}_\nu]
\eea
where $M_t$ is the mass of the top quark, $p_1$, $p_2$ are the momentum
of the incoming gluons and
\bea
&& g(\frac{4M_t^2}{(p_1+p_2)^2})~
=~[\sinh^{-1}(\frac{\sqrt{(p_1+p_2)^2}}{2M_t})]^2  \nonumber \\
&&{\rm if}~~~~~~~~(p_1+p_2)^2 \le 4M_t^2  \nonumber \\
&&~~~~~~~ \nonumber \\
&&~=-\frac{1}{4}[\ln(\frac{\sqrt{(p_1+p_2)^2}
+\sqrt{(p_1+p_2)^2-4M_t^2}}{|p_1+p_2|
-\sqrt{(p_1+p_2)^2-4M_t^2}})-i\pi]^2 \nonumber \\
&&{\rm if}~~~~~(p_1+p_2)^2 > 4M_t^2.
\label{gg}
\eea 

As the dominant radion ($\phi$) production mechanism at pp
collisions is given by the gluon fusion process 
\be
gg \rightarrow \phi
\ee
we will consider the gluon fusion process in this paper.
The partonic level cross section for the gluon fusion process 
is given by \cite{cheung}:
\bea
&&\hat{\sigma}_{gg \rightarrow \phi}(\hat s)~=~
\frac{\alpha_s^2 }{256 \pi \lambda_{\phi}^2}
\times |7+ \nonumber \\
&&\frac{4M_t^2}{\hat s}
[1+(1- \frac{4M_t^2}{\hat s})g(\frac{4M_t^2}{\hat s})]|^2
\eea
where
$g(\frac{4M_t^2}{\hat s})$ is given by eq. (\ref{gg}) with 
$\hat s =(p_1+p_2)^2$ being the partonic level center of mass energy. 
Convoluting with the gluon distribution function, the radion production
cross section in the central collision of two hadrons A and B is given by: 
\bea
&&\sigma_{AB \rightarrow \phi}(s)~=\int dx_1 \int dx_2 ~f_{g/A}(x_1,Q^2) 
\nonumber \\
&&f_{g/B}(x_2,Q^2)~ 
\delta(\hat s-M_{\phi}^2)~M_{\phi}^2~
\hat{\sigma}_{gg \rightarrow \phi}(\hat s)
\eea
where $\hat s=x_1 x_2 s$ with $\sqrt s$ being the total center of mass energy 
in the hadronic collisions and 
$f_{g/A}(x,Q^2)$ ($f_{g/B}(x,Q^2)$) are the gluon distribution functions
inside the hadron A(B). 
The factorization and re-normalization scale appearing in $f_{i/A}(x_1,Q^2)$
($f_{j/B}(x_2,Q^2)$) and $\alpha_s(Q^2)$ respectively,
is taken to be at $Q^2~=~M_{\phi}^2$.

For nuclear collisions at very high energy the calculation is similar to that
of the jet production in AA collisions at RHIC and LHC \cite{kj}. 
The parton distribution function inside a large nucleus is given by: 
\be
R_{a/A}(x_1, Q^2) = \frac{f_{a/A}(x_1,Q^2)}{A f_{a/N}(x_1,Q^2)}
\ee
where 
$f_{a/A}(x_1,Q^2)$ and $f_{a/N}(x_1,Q^2)$ are the parton distribution functions
inside the free nucleus and free nucleon respectively. The NMC and EMC
experiments show that $R_{a/A}(x_a,Q^2) \ne 1 $ for all values of $x$.
In fact there is a strong shadowing effect ($R_{a/A}(x_a,Q^2) < 1$)
for much smaller values of $x$ ($x << 0.01$). However, for TeV scale
physics the shadowing effects should not be important. 
This is because we may not probe the low $x$ physics as the scale $Q$
in our calculation is at least a TeV. For $Q$ = 1 TeV the average value
of Bjorken $x$ we probe is: $x_{av}=\frac{1}{5.5}$ which is very large. 
Even if we take the minimum transverse energy $E_{T,jet}$ or radion
mass $M_{\phi}$ to be 100 GeV in our calculation this corresponds to
the minimum average $x$ value $x_{av}^{min}=\frac{1}{55}$ where the 
shadowing effects
are not very large for $Q$= 100 GeV as can be seen from Fig. 1. In Fig. 1
we have used the EKS98 parametrization \cite{eks98} for gluon shadowing
effects at high energy. It can be seen that as the scale $Q$ becomes higher
the shadowing effects become weaker. The present shadowing analysis
\cite{eks98,kumano} do not cover the factorization scale $Q$ up to (and above)
1 TeV.
For this reason, and as shadowing effects are not important for TeV scale
physics we will use the unshadowed parton distribution function
$R_A(x,Q^2)$ = 1 in this paper. Similarly the saturation of gluons at 
LHC is not important for TeV scale physics as gluon saturation occurs at low
value of $x$ (equivalent to $Q$ $\sim$ 2 GeV \cite{al})
which is not covered in the TeV scale physics calculations.

\section{Results and Discussions}

In this section we will compute the cross section for graviton and
radion production at LHC both in pp and PbPb collisions at $\sqrt s^{NN}$= 14
and 5.5 TeV respectively. As the dominant
radion production cross section 
comes from $gg \rightarrow \phi$ fusion process which 
is a $2 \rightarrow 1$ process, 
the total cross section is finite and depend on the
mass of the radion. However, the total cross section for the graviton
production process: $ AB \rightarrow jet + graviton$ is a $2 \rightarrow 2$
process and is divergent at zero transverse momentum of the final state jet.
We present the results of graviton production total cross section above 
certain transverse momentum of the jet $p_{T,jet}=E_{T,jet}$. Rapidity range for
graviton production computation is taken to be $|y_{jet,G}| ~<~3$.
We use recent CTEQ6 \cite{cteq6}
parton structure function in our calculation. This structure function
covers the range of $Q$ up to 10 TeV.

In Fig. 2a we present the cross section for graviton + jet production
as a function of minimum transverse energy of the jet for a pp collision
at $\sqrt s$ = 14 TeV at LHC. As the dependence of cross section on number of 
extra dimensions in pp collisions at LHC is studied in more details
in \cite{wells,wells1} we will choose only one value of number of extra 
dimension (d=2) in case of the pp collision for the 
comparison purpose. All the curves in Fig. 2a
correspond to d=2. The minimum transverse energy of the jet is chosen
to be from 100 GeV to 2 TeV in our calculation.
The solid, dotted, dashed, dot-dashed 
lines correspond to Planck mass $M_P$= 2, 3, 4, 5 TeV respectively.
In Fig. 2b we present the cross section for graviton + jet production
as a function of minimum transverse energy of the jet for a PbPb collision
at $\sqrt s$ = 5.5 TeV at LHC. The solid and dotted lines correspond to
$M_P$ = 2 and 3 TeV respectively for number of extra dimensions d=2 
and the dashed and dot-dashed lines correspond to
$M_P$ = 2 and 3 TeV respectively but for d=4. 

A comparison with Fig. 2a for the case of pp collision 
for d=2 shows that the graviton production is
much larger in a PbPb collision than in a pp collision at LHC. 
Above $E_{T,jet}$= 100 GeV the graviton 
production in a PbPb collision is 2220 times larger than that in a pp
collision at LHC. This number is independent of the Planck mass
as Planck mass cancel in the ratio. This can be seen in Fig. 4a 
where the ratio of total cross section in a PbPb collision to a pp
collision at LHC is plotted as a function of minimum transverse
energy of the jet $E_{T,jet}$. Above 
$E_{T,jet}$= 500, 1000, 1500 GeV this ratio is 400, 43, 2 respectively
in case of the number of extra dimension d=2.
For d=4 this ratio is 470, 100, 10, 0.5 above the minimum transverse
energy of the jet 
$E_{T,jet}$ = 100, 500, 1000, 1500 GeV which can also be seen from Fig. 4a.
One can observe that the graviton production is much more enhanced in a
PbPb collision than in a pp collision for minimum transverse energy
of the jet upto $E_{T,jet}$ = 1.5 TeV.
Hence PbPb collisions at LHC can be considered as graviton production 
factory for minimum transverse energy of the jet up to $E_{T,jet}$= 1.5 TeV 
and pp collisions at LHC can be considered as graviton production 
factory for minimum transverse energy of the jet above $E_{T,jet}$= 1.5 TeV.
Hence pp and PbPb collisions at LHC will provide the best oppurtinties
to find the possible existence of extra dimensions. 

In Fig. 3a we present the radion production cross section from the gluon
fusion process in a pp collision at LHC as a function of radion mass
$M_{\phi}$. Different lines are for different values of $\lambda_{\phi}$,
the strength of the radion coupling to the standard model particles.
The solid, dashed, dotted lines correspond to $\lambda_{\phi}$=1, 2, 3
TeV respectively. The radion production is described in a 5-d theory
hence there is no extra dimension dependence on the cross section as
was in the case of graviton production. In Fig. 3b we present the 
radion production cross section from the gluon
fusion process in a PbPb collision at LHC as a function of radion mass
$M_{\phi}$. 
The solid, dashed, dotted lines correspond to $\lambda_{\phi}$=1, 2, 3
TeV respectively. 

As can be seen from Fig. 3 the radion production in a PbPb collision is much
higher than in a pp collision at LHC. For radion of mass $M_{\phi}$ = 100 GeV
the radion production cross section in a PbPb collision is 1650 times larger
than that in a pp collision at LHC. In Fig. 4b we present the ratio of the
radion production cross section in a PbPb collision to that in a pp collision at
LHC as a function of radion mass. The ratio is 1650, 460, 180, 67, 23 and 1 for
radion of masses 100, 500, 1000, 1500, 2000 and 3000 GeV
respectively. This ratio is independent
of radion coupling strength $\lambda_{\phi}$. It can be observed that the
radion cross section is much larger in a PbPb collision than in a pp
collision even up to radion mass of 3 TeV. Therefore PbPb collisions at
LHC can be considered as radion production factory for radion  
masses up to 3 TeV and pp collisions at LHC can be considered as 
radion production factory for radion masses above 3 TeV. 

\section{Conclusion}

If Planck scale is around a TeV in theories with large extra dimension, 
then gravitons and radions should be
produced in high energy hadronic collisions at the LHC. In this paper we
have computed the direct graviton and radion production cross sections in a 
pp and PbPb collision at LHC at $\sqrt s^{NN}$= 14 and 5.5 TeV respectively.
The direct graviton (plus jet) production cross section
in a PbPb collision is enhanced in comparison to a pp collision at LHC
depending upon the minimum transverse energy $E_{T,jet}^{min}$ of the jet.
For two extra dimensions and above $E_{T,jet}^{min}$ = 100, 500, 1000, 1500 GeV 
the ratio of graviton production cross section in a PbPb collision
to that in a pp collision is found to be
2220, 400, 43, 2. For four extra dimensions 
this ratio is found to be 470, 100, 10, 0.5. The radion production
cross section is also found to be enhanced in a PbPb collision than
in a pp collision at LHC. The ratio of the radion production cross section
in a PbPb collision to that in a pp collision at LHC is found to be
1650, 460, 180, 67, 23 and 1 for radion of mass 100, 500, 1000, 1500, 2000
and 3000 GeV respectively. Hence PbPb (pp) collisions at LHC can be 
considered as graviton production factory up to (above)
the minimum transverse momentum of the jet $\sim$ 1.5 TeV.
Similarly PbPb (pp) collisions at
LHC can be considered as radion production factory for radion 
masses up to (above) 3 TeV.
Hence pp and PbPb collisions at LHC will provide the best oppurtinties
to find the possible existence of extra dimensions.

\acknowledgements
I am thankful to John Terning for introducing me to this problem. I also
thank him for discussions throughout this work and for careful reading
of the manuscript. I would like to acknowledge duscussions
with Andrew Chamblin, Fred Cooper, Joshua Erlich, Stavros Mouslopoulos 
and Yuri Shirman.

\end{multicols} 

\vspace{4cm}

\parbox{15cm}{
\parbox[t]{15cm}
{\begin{center}
\mbox{\epsfxsize=6.5cm\epsfysize=4cm\epsfbox{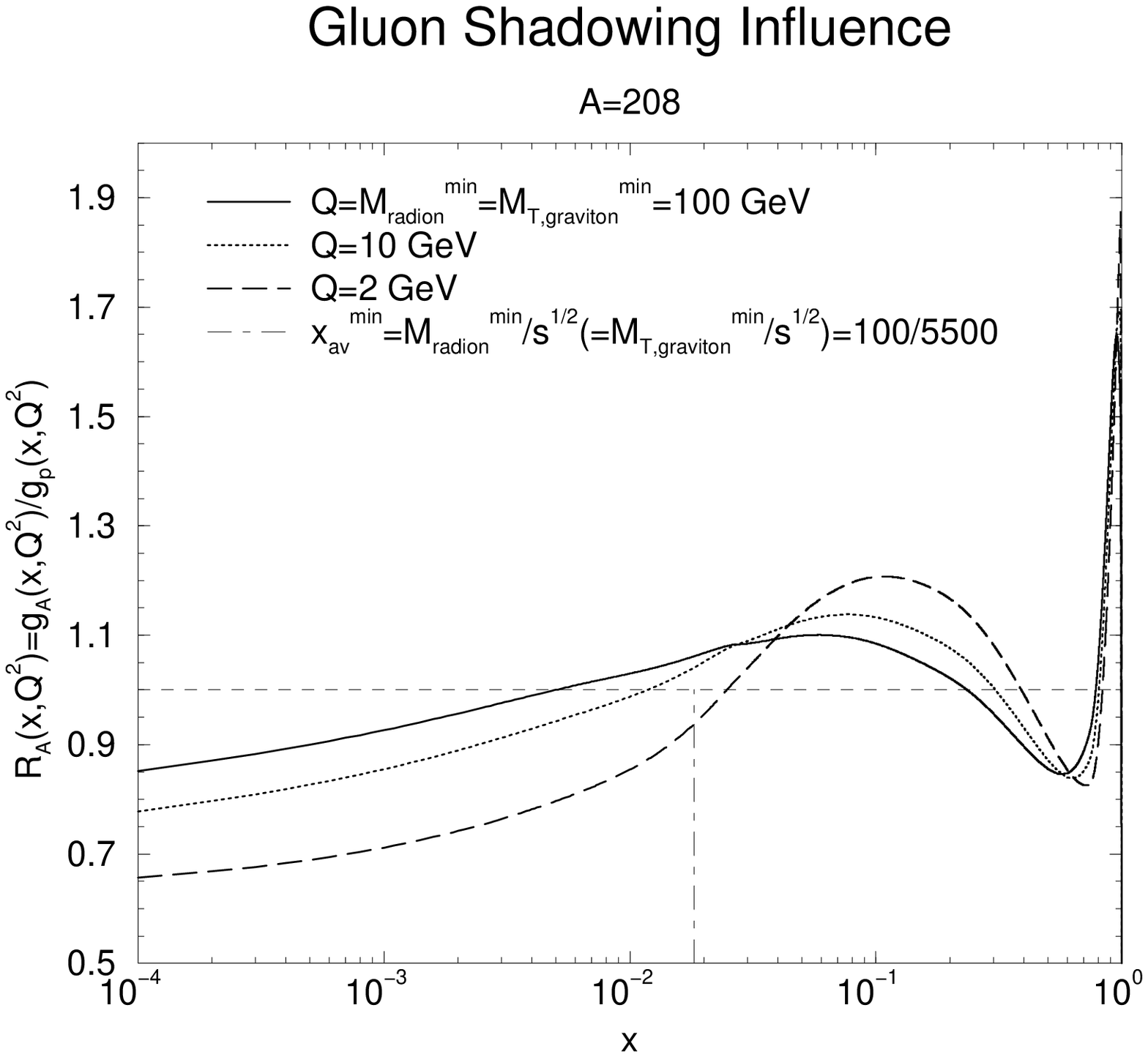}}
\end{center}}
~~~~~~~{{\small FIG. 1: The gluon shadowing at LHC and TeV scale physics.
There is no significant impact of gluon shadowing on TeV scale physics
as TeV scale physics do not correspond to small Bjorken $x$.
}}}

\vspace{5cm}

\parbox{15cm}{
\parbox[t]{7cm}
{\begin{center}
\mbox{\epsfxsize=6.5cm\epsfysize=4cm\epsfbox{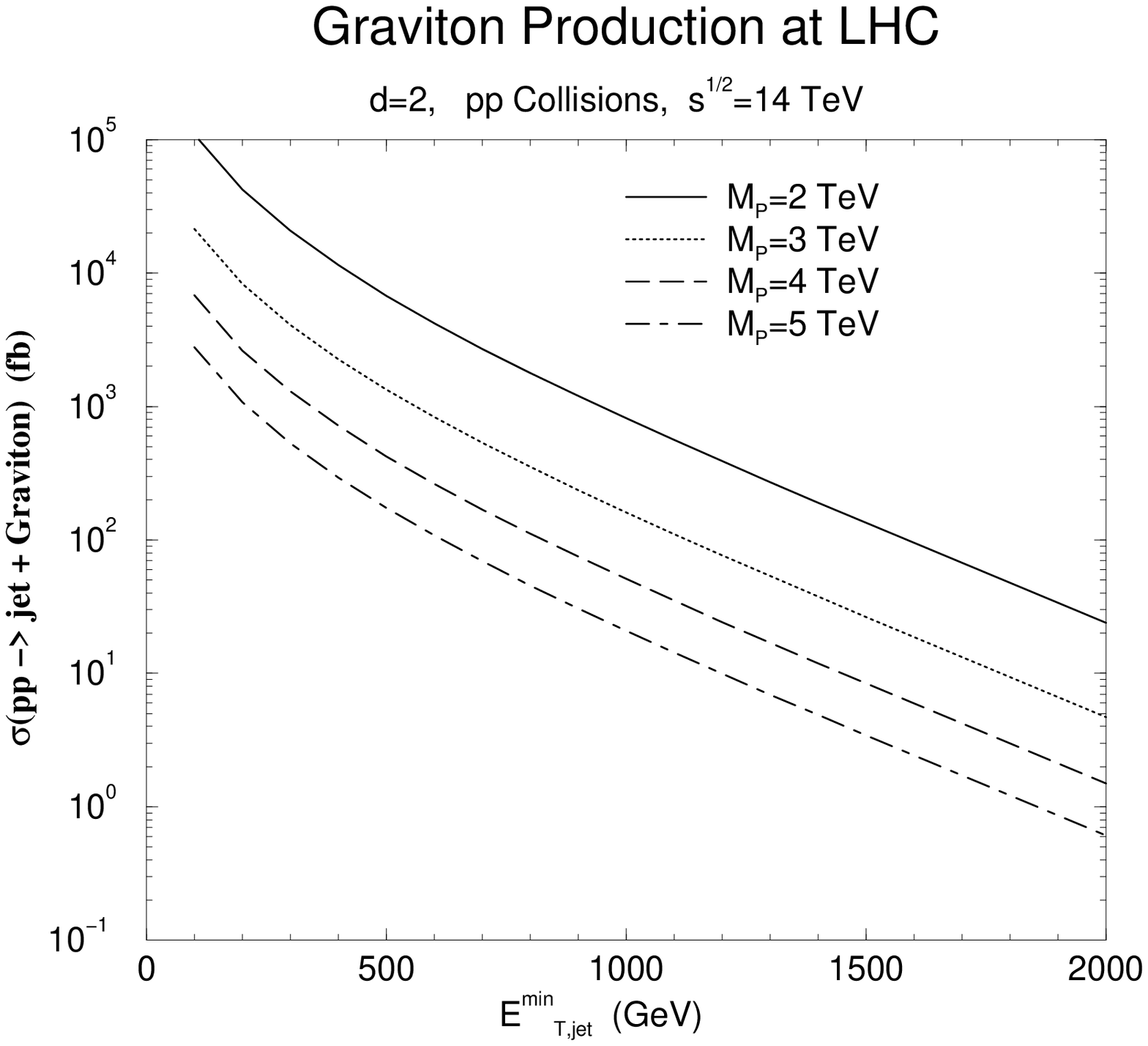}}
\end{center}}
\hspace{.5cm}\parbox[t]{7cm}
{\begin{center}
\mbox{\epsfxsize=6.5cm\epsfysize=4cm\epsfbox{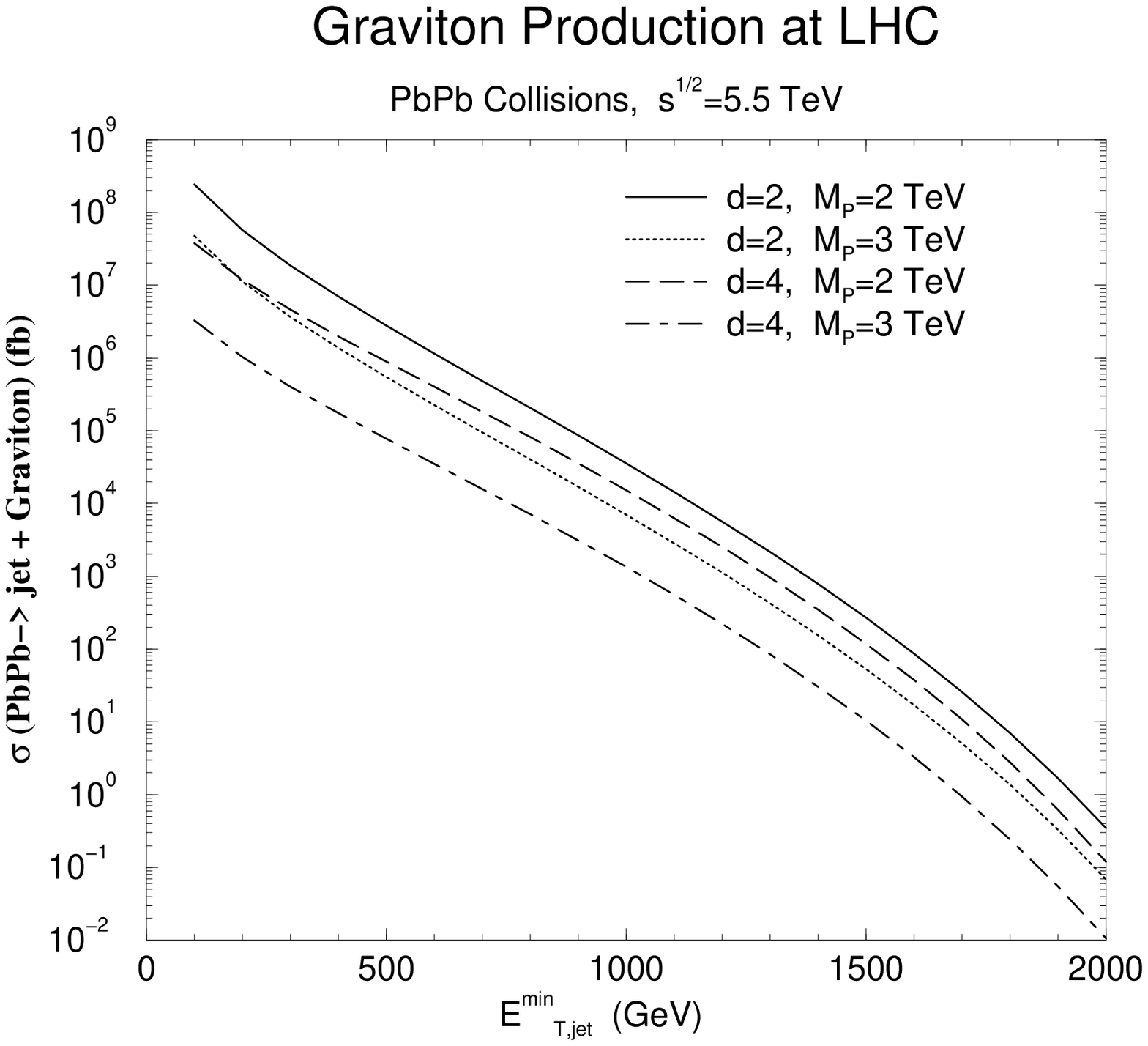}}
\end{center}}
\parbox[t]{7cm}
{{\small FIG. 2a: The total cross section for graviton production
in a pp collision at $\sqrt s^{NN}$ = 14 TeV at LHC as a function
of minimum transverse energy of the jet.
}}
\hspace{0.8cm}\parbox[t]{7cm}
{{\small FIG. 2b:
The total cross section for graviton production
in a PbPb collision at $\sqrt s^{NN}$ = 5.5 TeV at LHC as a function
of minimum transverse energy of the jet.
}}}

\vspace{3cm}

\parbox{15cm}{
\parbox[t]{7cm}
{\begin{center}
\mbox{\epsfxsize=6.5cm\epsfysize=4cm\epsfbox{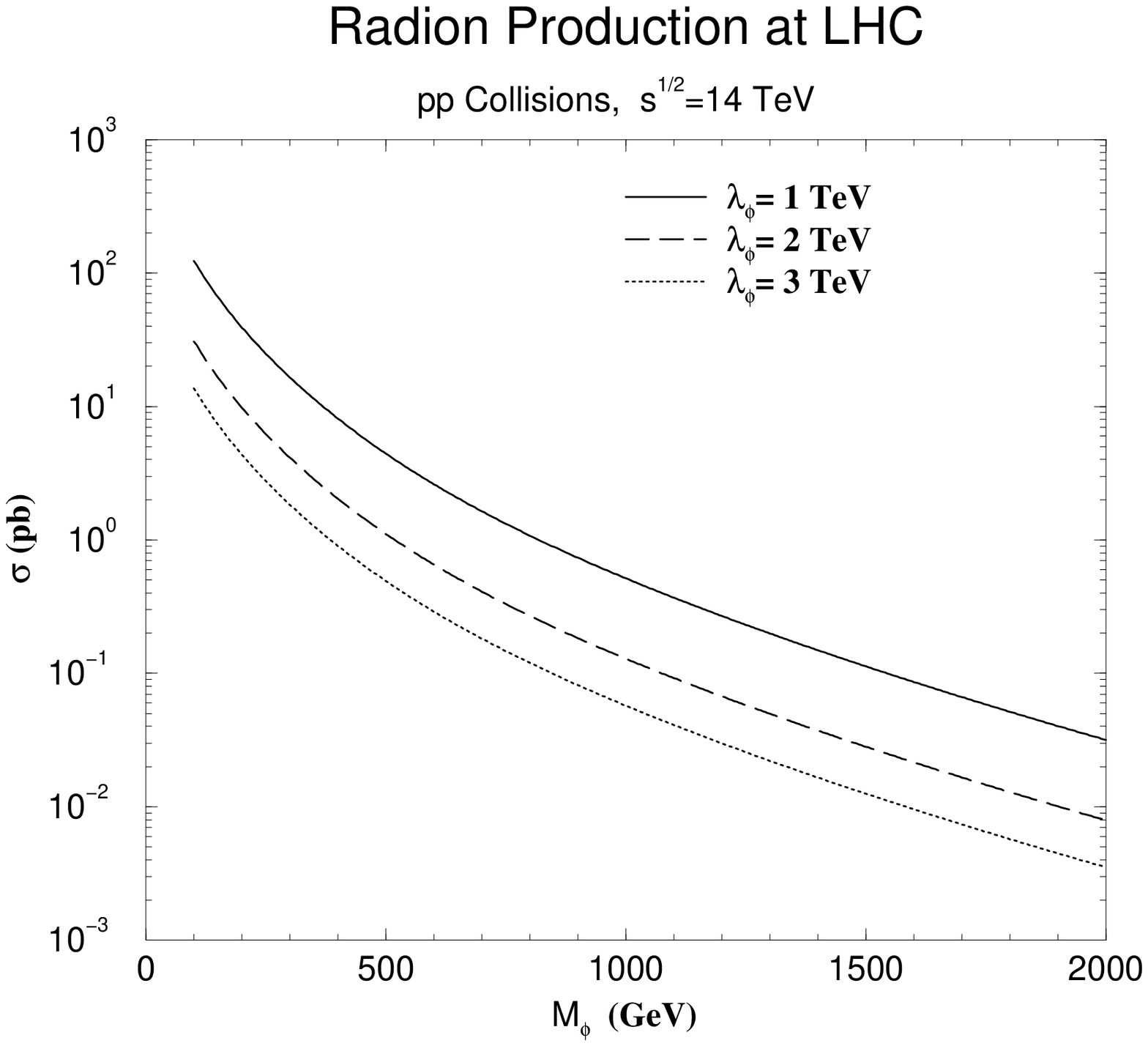}}
\end{center}}
\hspace{.5cm}\parbox[t]{7cm}
{\begin{center}
\mbox{\epsfxsize=6.5cm\epsfysize=4cm\epsfbox{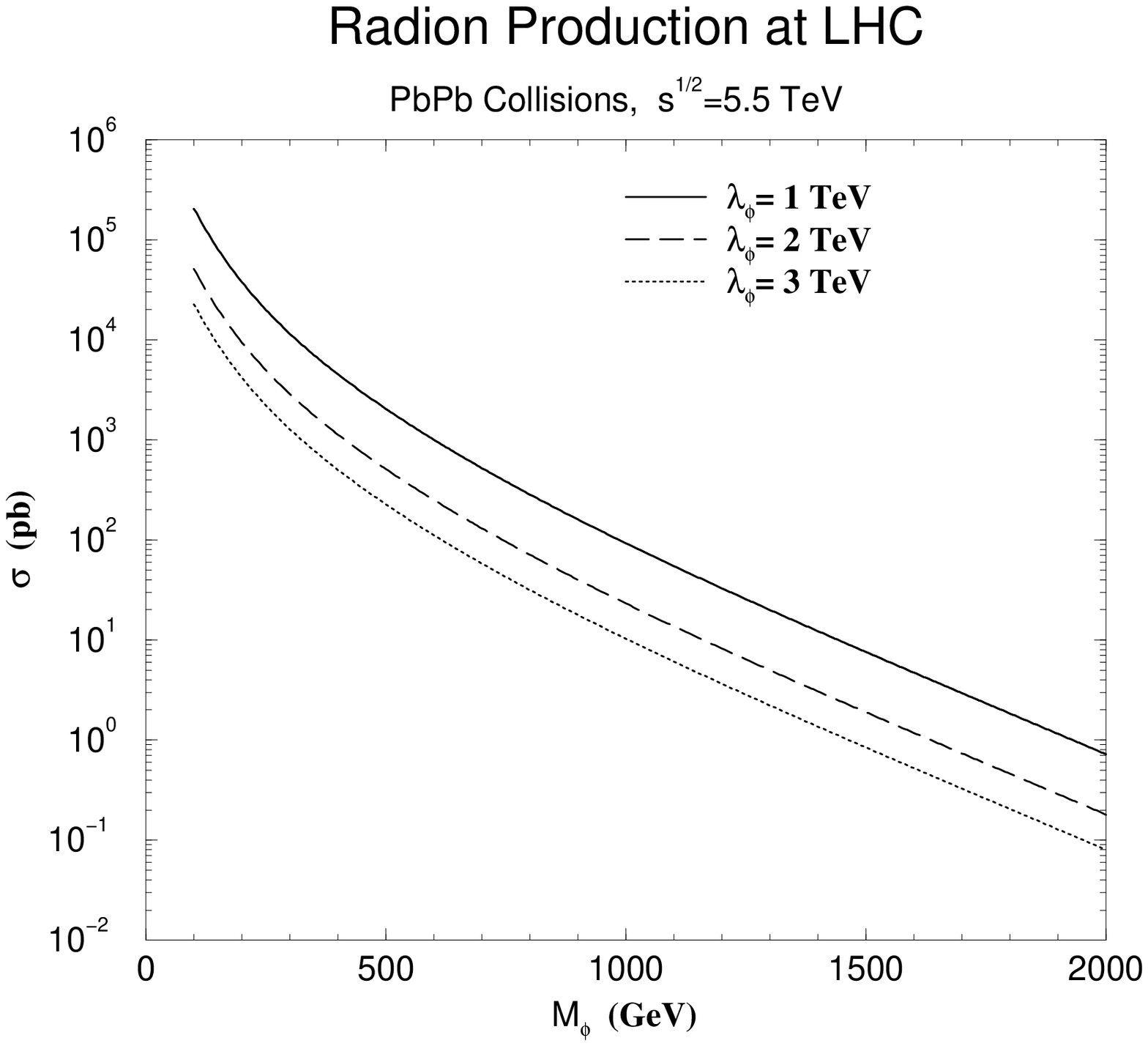}}
\end{center}}
\parbox[t]{7cm}
{{\small FIG. 3a: The total cross section for radion production
in a pp collision at $\sqrt s^{NN}$ = 14 TeV at LHC as a function
of radion mass from the gluon fusion process.
}}
\hspace{0.8cm}\parbox[t]{7cm}
{{\small FIG. 3b:
The total cross section for radion production
in a PbPb collision at $\sqrt s^{NN}$ = 5.5 TeV at LHC as a function
of radion mass from the gluon fusion process.
}}}

\vspace{5cm}

\parbox{15cm}{
\parbox[t]{7cm}
{\begin{center}
\mbox{\epsfxsize=6.5cm\epsfysize=4cm\epsfbox{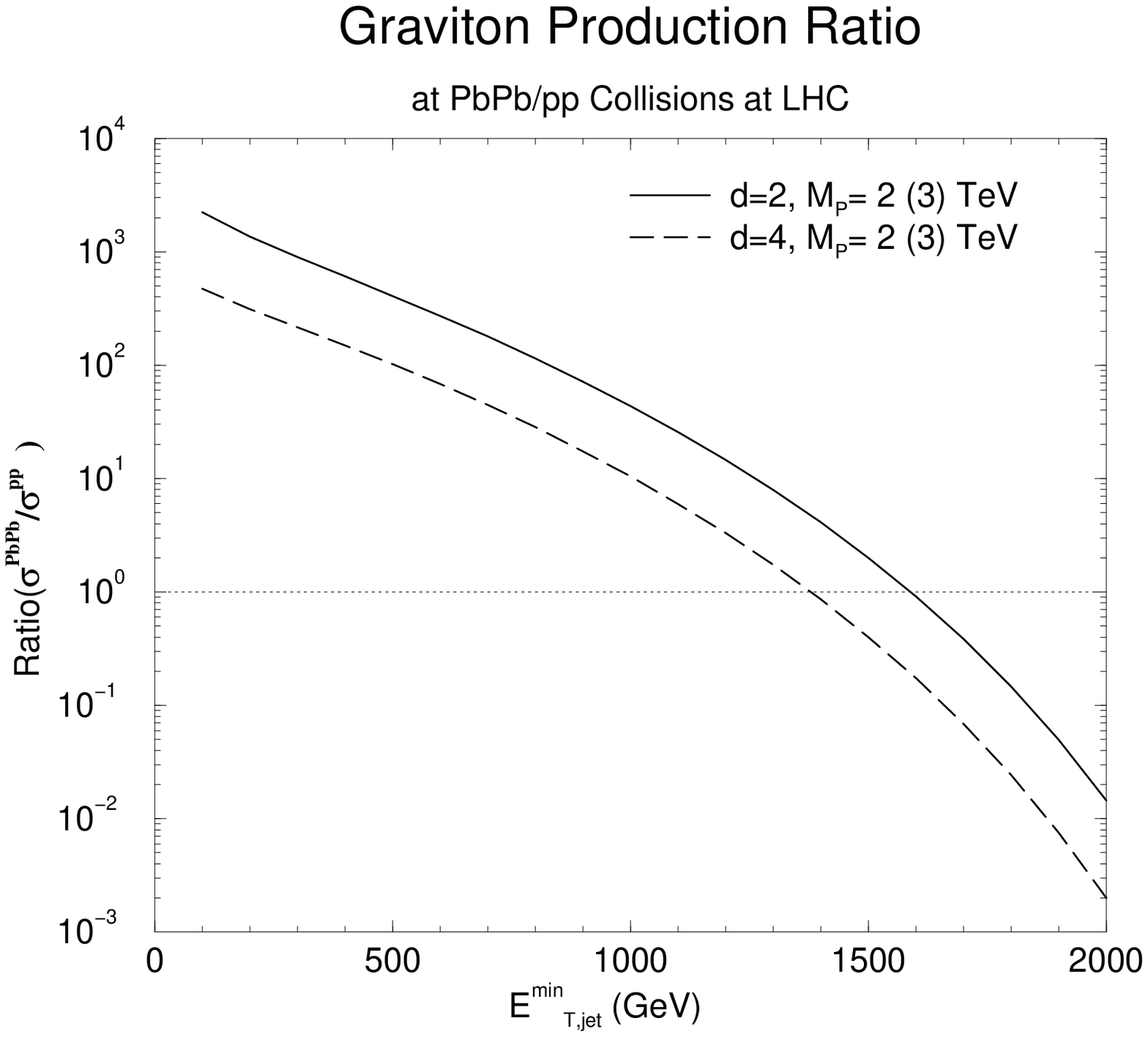}}
\end{center}}
\hspace{.5cm}\parbox[t]{7cm}
{\begin{center}
\mbox{\epsfxsize=6.5cm\epsfysize=4cm\epsfbox{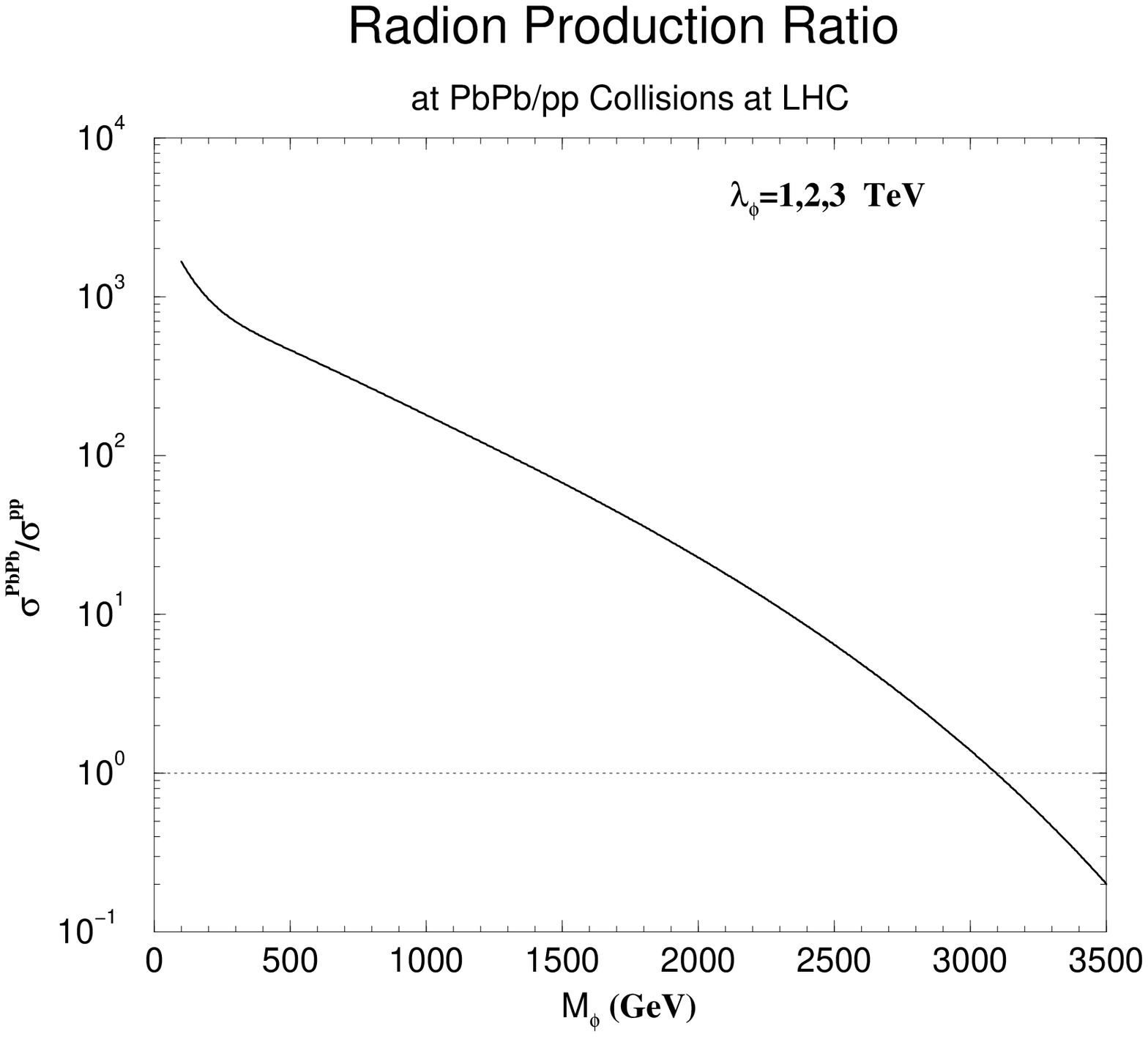}}
\end{center}}
\parbox[t]{7cm}
{{\small FIG. 4a: The ratio of the 
total cross section for gravtion production
in a PbPb collision at $\sqrt s^{NN}$ = 5.5 TeV to a pp collision
at $\sqrt s$= 14 TeV at LHC as a function
of minimum transverse energy of the jet.
}}
\hspace{0.8cm}\parbox[t]{7cm}
{{\small FIG. 4b:
The ratio of the total cross section for radion production
in a PbPb collision at $\sqrt s^{NN}$ = 5.5 TeV to a pp collision
at $\sqrt s$= 14 TeV at LHC as a function
of radion mass from the gluon fusion process.
}}}

\end{document}